# Guiding of positive streamers in nitrogen, argon and $N_2$-$O_2$ mixtures by very low $n_e$ laser-induced pre-ionization trails

S. Nijdam[1], E. Takahashi[2]

[1] Eindhoven University of Technology, Dept. Applied Physics P.O. Box 513, 5600 MB Eindhoven, The Netherlands
[2] National Institute of Advanced Industrial Science and Technology (AIST),1-2-1 Namiki, Tsukuba Ibaraki 305-8564, Japan

In previous work we have shown that positive streamers in pure nitrogen can be guided by a laser-induced trail of low electron density. Here we show more detailed results from such measurements. We show the sensitivity of this laser-guiding on pressure $p$ and found that the maximum delay between the laser pulse and voltage pulse for guiding scales with something between $1/p$ and $1/p^2$. We also show that when we use a narrower laser beam the laser guiding occurs less frequent and that when we move the laser beam away from the symmetry axis, guiding hardly is observed. Finally we show that laser guiding can also occur in pure argon.

## 1. Introduction

Positive streamers are known to be sensitive for variations in the background ionization density, or more specifically the electron density. Because of their positive charge, such streamers require free electrons in front of them in order to propagate. These electrons can either come from photo-ionization due to the streamers themselves, or already be present in the gas, created by a variety of sources. In case of streamers in air, the photo-ionization mechanism including nitrogen and oxygen is very efficient, leading to a very high electron density in front of the streamer.

We previously reported laser-guided streamers in pure nitrogen and nitrogen-oxygen mixtures [1]. Here it was found that in mixtures with less than 0.5% oxygen, laser guiding of streamers would occur under our experimental conditions: 133 mbar of gas in a 130 mm point-plane configuration driven by 4 – 10 kV positive pulses with laser induced ionization trails with electron densities of $10^8$ – $10^9$ cm$^{-3}$. Such very weakly ionized trails do not lead to any significant modification of the local electric field, but they do guide streamers, even in directions almost perpendicular to the background field. It was also shown that these experimental observations were backed by particle simulations, although for somewhat different conditions, limited by computational expenses.

Here we will look at this laser guiding in more detail, in order to better understand the streamer propagation mechanism in general. We will also show streamer guiding in pure argon which has not been reported before.

## 2. Experimental set-up and methods

The experimental set-up and methods are identical to the ones described in [1]. The set-up consists of a vacuum vessel containing a tip – plate geometry, combined with a laser, a camera system and a pulsed high voltage source. The most important components are shown in figure 1.

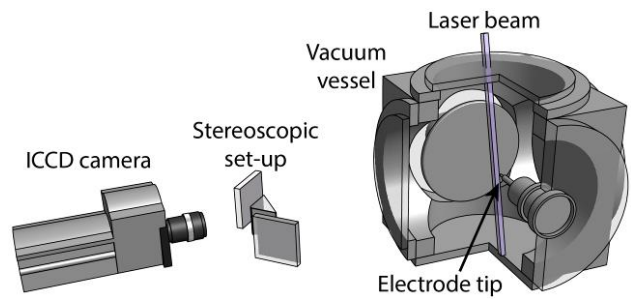

Figure 1: Overview of the experimental set-up with indication of the laser beam path and the stereoscopic camera set-up. Taken from [1].

The laser beam is produced by a KrF excimer laser which makes pulses at 248 nm with about 1 mJ energy and 20 ns length per pulse. It is shaped by shutters into a rectangular, unfocused, beam with varying dimensions. Some micro- to milliseconds after the laser pulse, a positive high voltage pulse is applied to the electrode tip with an amplitude of 4 – 10 kV, a length of 0.6 – 1.5 µs and a rise/fall time of about 15 ns.

The vacuum vessel is filled with various gases at pressures between 25 and 250 mbar. The resulting streamer discharges are imaged with an ICCD camera behind a simple stereoscopic set-up (see [2]), enabling a complete 3D reconstruction of the location of streamer channels with respect to the laser beam location.

## 3. Results and discussion
### 3.1. Laser guiding as function of pressure

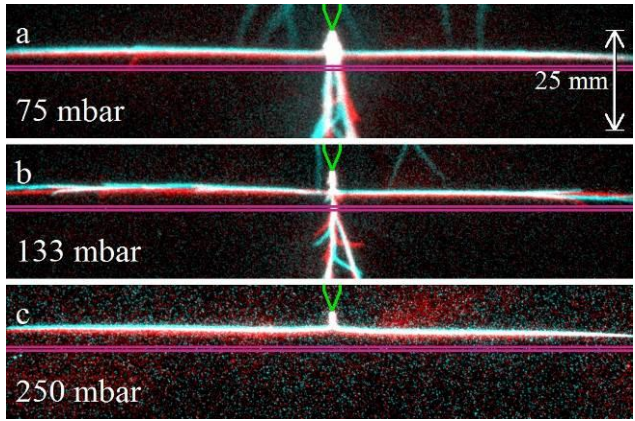

Figure 2: Laser guided streamers in pure nitrogen at different pressures. Measured with voltage pulses of 6 kV magnitude and 1 μs duration. The laser beam was 1 mm high and 9 mm wide.

In figure 2 we show laser guiding of streamers in a pure nitrogen atmosphere for three different pressures. It can be observed that the guiding is independent of pressure within our parameter range. Also the offset between the laser sheet and the lowest guided streamer is almost independent of pressure under these conditions.

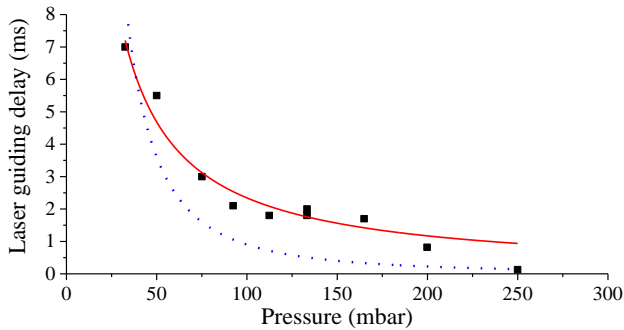

Figure 3: Maximum delay between laser and high voltage pulse for which streamer guiding is observed as function of pressure. Fit lines represent $1/p$ (red, solid) and $1/p^2$ (blue, dotted).

We have measured the maximum delay between the laser pulse and the voltage pulse for which guiding is still observed as function of pressure, see figure 3. Over this large pressure range it was not possible to keep the voltage constant and still produce similar looking streamers. Therefore voltage was varied from 5.2 kV to 8.3 kV (higher voltages were used with higher pressures).

It can be observed that this delay scales roughly with the inverse of the pressure but could also be fitted with the inverse of the pressure squared. As was shown in [3], most relevant electron loss processes scale with $1/n^2$ where $n$ is the gas density. From this one would expect scaling with $1/p^2$. However, the required (critical) electron density for laser guiding will also depend on gas density.

### 3.2. Laser guiding with thin and/or offset laser beams

When changing the 9 mm wide laser beam to a smaller, 1 mm wide laser beam the laser guiding is still quite similar to the wide beam, as is shown in figures 3a and 3c. However, the number of non-guided streamers is higher for this thin laser beam than for the 9 mm wide laser beam. When this thin laser beam is also moved away from the symmetry axis of the vacuum vessel and the electrode tip, the number of guided streamers decreases dramatically. For the 1x1 mm$^2$ laser beam that is 4 mm off-axis (figure 3b) no guiding at all was observed, while for the 9x1 mm$^2$ beam at the same position guiding was only observed in three out of twenty discharges.

This indicates that the streamers are not attracted to the increased electron density in the laser trail. The streamers do follow this trail when they encounter it, but do not move towards it.

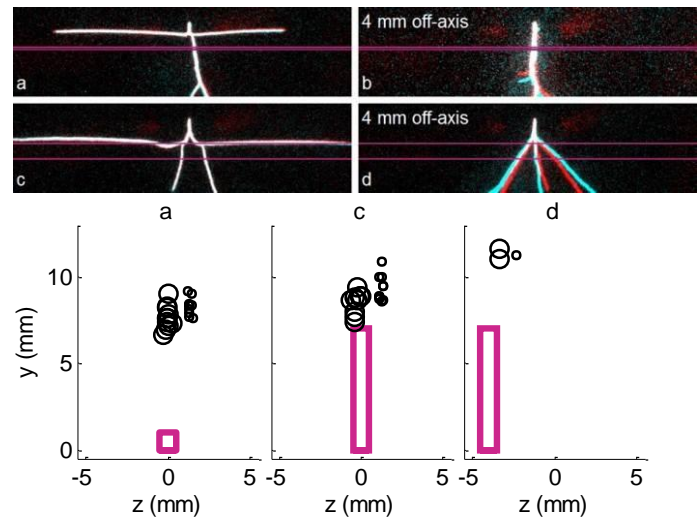

Figure 4: Laser guided streamers in pure nitrogen for various thin beam shapes and offsets. Measured in 133 mbar pure nitrogen with voltage pulses of 4.8 kV magnitude and 1.3 μs duration. The bottom graphs represent the position of guided streamers with respect to the laser beam for the conditions of images a, c and d. Positions are obtained from twenty discharges per beam shape/position. For the beam position shown in b no guided streamers were observed.

### 3.3. Guiding in argon

Besides nitrogen-oxygen mixtures, we have also studied the effect of laser guiding in pure argon. We found that in pure argon guiding of streamers occurs in a similar way as in pure nitrogen. Images of guided streamers in argon are plotted in figure 5. Although in general very similar to guiding in pure nitrogen, there is one important difference: In argon it seems that any streamer that crosses the laser beam, whether it is guided or not, becomes significantly thicker and branches less. This was previously observed by Takahashi *et al*. [4]. Probably because of this extra thickness of guided streamers, the occurrence of multiple guided streamers above each other is much rarer in argon than in nitrogen. The maximum delay between laser and voltage pulse with guiding is not so different between argon and nitrogen, both are in the millisecond range for the conditions under investigation.

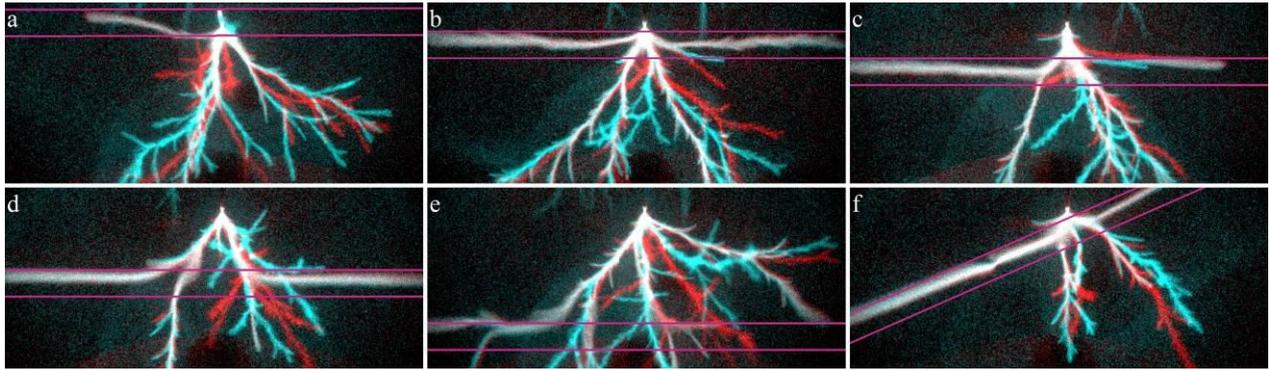

Figure 5: Laser guided streamers in 133 mbar pure argon for different laser positions. Measured with 4.4 kV voltage pulses 1.2 μs duration. The laser beam was 10 mm high and 9 mm wide.

### 4. Conclusions

We have shown that laser guiding of streamers can occur under a wide variety of conditions in pure nitrogen and argon at pressures between 33 and 250 mbar. The maximum delay between laser and voltage pulse scales with inverse or inverse squared of pressure, most likely due to the loss processes of the laser-induced free electrons.

When reducing the width of the laser beam, the efficiency of laser guiding slightly reduces. This efficiency further reduces, to almost zero, when this thin beam is moved a few millimetres out of the plane of symmetry. From this we can conclude that the streamers are not attracted to the laser induced trail and therefore that the laser guiding is not due to electrostatic attraction of the streamers.

The fact that laser guiding also is present in pure argon shows that it is not related to any specific property of nitrogen. In previous work [1] we have shown that guiding (under our conditions) does not occur in air, which we attributed to photo-ionization that makes the streamers less sensitive to local increases in background electron density. In this perspective it should not be surprising that guiding is possible in pure argon as photo-ionization is not expected to play a big role in this gas.